\documentclass[prb,amssymb,twocolumn,showpacs,groupedaddress,superscriptaddress,floatfix]{revtex4}

\usepackage{graphicx}
\usepackage{times}
\usepackage{psfrag}

\newcommand{\be}{\begin{equation}}
\newcommand{\ee}{\end{equation}}
\newcommand{\bae}{\begin{eqnarray}}
\newcommand{\eae}{\end{eqnarray}}

\def\CC{{\rm\kern.24em \vrule width.04em height1.46ex depth-.07ex
    \kern-.30em C}}
\def\P{{\rm I\kern-.25em P}}

\def\bbbc{{\mathchoice {\setbox0=\hbox{$\displaystyle\rm C$}\hbox{\hbox
to0pt{\kern0.4\wd0\vrule height0.9\ht0\hss}\box0}}
{\setbox0=\hbox{$\textstyle\rm C$}\hbox{\hbox
to0pt{\kern0.4\wd0\vrule height0.9\ht0\hss}\box0}}
{\setbox0=\hbox{$\scriptstyle\rm C$}\hbox{\hbox
to0pt{\kern0.4\wd0\vrule height0.9\ht0\hss}\box0}}
{\setbox0=\hbox{$\scriptscriptstyle\rm C$}\hbox{\hbox
to0pt{\kern0.4\wd0\vrule height0.9\ht0\hss}\box0}}}}

\def\bbbz{{\mathchoice {\hbox{$\sf\textstyle Z\kern-0.4em Z$}}
{\hbox{$\sf\textstyle Z\kern-0.4em Z$}}
{\hbox{$\sf\scriptstyle Z\kern-0.3em Z$}}
{\hbox{$\sf\scriptscriptstyle Z\kern-0.2em Z$}}}}

\begin{document}

\title{Single-site entanglement at ``superconductor"-insulator transition in the Hirsch model}

\author{Alberto Anfossi}
\affiliation{Dipartimento di Fisica
del Politecnico, C.so Duca degli Abruzzi, 24, I-10129, Torino, Italy}
\author{Cristian \surname{Degli Esposti Boschi}}
\affiliation{CNR-INFM, Unit\`a di Ricerca di Bologna, V.le Berti-Pichat, 6/2, I-40127, Bologna, Italy}
\author{Arianna Montorsi}
\affiliation{Dipartimento di Fisica
del Politecnico, C.so Duca degli Abruzzi, 24, I-10129, Torino, Italy}
\author{Fabio Ortolani}
\affiliation{Dipartimento di Fisica dell'Universit\`a di Bologna and INFN, Via Irnerio, 46, I-40126, Bologna, Italy}
\affiliation{CNR-INFM, Unit\`a di Ricerca di Bologna, V.le Berti-Pichat, 6/2, I-40127, Bologna, Italy}

\date{Received 9 November 2005}

\begin{abstract}
We investigate the transition to the insulating state in the one-dimensional Hubbard model with bond-charge interaction $x$ (Hirsch model), at half-filling and $T=0$. By means of the density-matrix renormalization group algorithm the charge gap closure is examined by both  standard finite-size scaling analysis and looking at singularities in the derivatives of single-site entanglement. The results of the two techniques show that a quantum phase transition takes place at a finite Coulomb interaction $u_{\rm c}(x)$ for $x\gtrsim 0.5$. The region $0<u<u_{\rm c}$ turns out to have a superconducting nature, at least for not too large $x>x_{\rm c}$.
\end{abstract}


\maketitle

\section{Introduction}

Recently \cite{HIR,SCH} the role of the bond-charge interaction $x$ in inducing interesting physics in low-dimensional correlated electron systems has been underlined. Such an interaction originates from the Coulomb repulsion between electrons in a solid, and its actual value is in general smaller than the on-site Coulomb interaction (Hubbard $u$ term), though not negligible in many cases. Moreover, the $x$ term breaks the particle-hole symmetry of the model, which feature is known to be relevant in relation to high-$T_{\rm c}$ superconductivity. In particular, in the one-dimensional case, the bond-charge interaction has been proven to play a relevant role in real materials. \cite{CBM}

A complete study of the behavior of a one-dimensional Hubbard system in which the $x$ term is included is still missing, though a certain number of interesting and competing information are known. First of all, $x=0$ corresponds to the standard Hubbard model, in which case at half-filling ($n=1$) and $T=0$ a metal-insulator transition takes place at $u_{\rm c}=0$, the insulating state being characterized by antiferromagnetic order. Moreover, bosonization studies \cite{JAP} show that the critical value $u_{\rm c}$ is independent from the actual value of the bond-charge interaction, at least for $x\ll 1$. On the other hand, an exact solution of the $x=1$ case is also available. \cite{AAS} There the charge gap responsible for the insulating phase at $n=1$ opens at $u_{\rm c}=4 $. The nonvanishing value of $u_{\rm c}$ is a clear indication of the failure of the bosonization approximation in this regime. Still, at $x=1$ and for $ u \leq u_{\rm c}(n) $, the system enters a phase characterized by off-diagonal long-range order (ODLRO \cite{YANG}) and nonvanishing pairing correlation, both features being an indication of possible superconducting order, which is absent at $x=1$ due to the high degeneracy of the ground state (g.s.). The possible presence of superconducting order has been investigated also for $x\neq 1$ at various $n$. It turns out that this is the case, at least for low enough $u$ and appropriate $x$ and $n$ values; \cite{AAG,QUSH,BUL} also, a spin gap opens for $n>1$, \cite{SCH,QUSH} whereas, as for the transition to the insulating phase, results at $x\neq 1$ are far from exhaustive, \cite{BUL,AGAH} though supporting the possibility of a finite $u_{\rm c}$ away from the weak-coupling limit.

To resume the physical picture emerging from previous studies, at $n=1$ and $T=0$ it is expected that the system undergoes a change at intermediate values of $x$, passing from a situation in which a metal-antiferromagnetic insulator transition takes place at vanishing $u$, to a situation in which a possible superconductor-insulator transition occurs at finite $u$.

The purpose of the present paper is to investigate such a change. This is achieved by means of a composite analysis of the numerical results obtained using the density-matrix renormalization group (DMRG) algorithm (Sec. III). On the one hand, the study of charge gap closure is performed by means of its finite-size scaling (FSS) behavior (Sec. III A). On the other hand, quite recent results \cite{AGMT} have proven that, at least at $x=1$, the quantum phase transition (QPT) to the insulating state at half-filling corresponds to a divergence (in the thermodynamic limit) in the partial derivative with respect to $u$ of the single-site entropy of entanglement. \cite{ZAN} $\mathcal{S}$ Accordingly, we investigated the behavior of $\mathcal{S}$ also for $x\neq 1$ (Sec. III B). Furthermore, in Sec. III C both spin-gap and pair-correlation behaviors are analyzed within the noninsulating phase.

\section{The Hirsch model}

The model we deal with is the bond-charge extended Hubbard model first introduced by Hirsch. \cite{HIR2} It is described by the Hamiltonian
\be\label{ham_bc}
    \mathcal{H} = -\sum_{\langle i,j\rangle\sigma} [1 - x (n_{i\bar{\sigma}}+n_{j \bar{\sigma}})]c_{i \sigma}^\dagger c_{j\sigma}+ u \sum_i n_{i \uparrow}n_{i \downarrow}  \; ,
\ee
where $c_{{i} \sigma}^\dagger$ and $c_{{i}\sigma}$ are fermionic creation and annihilation operators on a one-dimensional chain of length $L$; $\sigma = \uparrow,\downarrow$ is the spin label, $\bar{\sigma}$ denotes its opposite, ${n}^{}_{j \sigma} = c_{j \sigma}^\dagger c_{j \sigma}^{}$ is the electron charge with spin $\sigma$, and $\langle {i} , \, {j} \rangle$ stands for neighboring sites; $u$ and $x$ are the (dimensionless) on-site Coulomb repulsion and bond-charge interaction parameters, respectively.

The model Hamiltonian (\ref{ham_bc}) has su($2)\oplus$u$(1)$ symmetry, since it commutes with total spin and charge operators; correspondingly, their eigenvalues $S^z_{\rm tot}$ and $N$ are good quantum numbers. Due particle-hole transformation properties, the parameter region $0\leq x\leq 1$ is representative of all $x$ values.

In the following we shall focus on the half-filled regime. For $x=1$ $\mathcal{H}$ acquires an extra symmetry, since in this case the so-called $\eta$-pairing operator $\eta= \sum_j \eta_j\doteq\sum_j c_{j\downarrow} c_{j\uparrow}$ and its hermitian conjugate both commute with the Hamiltonian, which also preserves the total number of doubly occupied sites. These symmetries can be exploited to obtain the complete spectrum of the model, from which both zero-temperature \cite{AAS} and finite-temperature \cite{DOMO} properties can be gained. In particular the g.s. phase diagram is characterized by two QPT's. At $u=4$ a charge gap opens and the system enters an insulating phase, whereas for $-4\leq u\leq 4$ a gapless phase characterized by superconducting correlations and ODLRO is approached. The latter reads $\lim_{|i-j|\rightarrow\infty} \langle\eta_i^\dagger \eta_j\rangle = n_d^2$, with $n_d$ the average number of doubly occupied sites ($0\leq n_d \leq 1/2$). For $u < -4$ a further phase appears, which contains just empty and doubly occupied sites, and is again characterized by ODLRO. Noticeably, all phases containing singly occupied sites are highly degenerate, since the spin orientation is arbitrary. The $x=1$ case has also been explored \cite{AGMT} by means of single-site entanglement. In the sector with zero magnetization ($S^z_{\rm tot}=0$) the latter reads
\be\label{le}
    \mathcal{S}=-\left[ 2 n_d \log_2 n_d+(1-2 n_d)\log_2\left(\frac{1}{2} - n_d\right)\right]\;.
\ee
In the thermodynamic limit the above QPT's appear as divergencies in $\partial_u \mathcal{S}$. In particular, the transition to the insulating state is signaled by an algebraic divergence of the latter quantity. $\mathcal{S}$ has been studied at $x=0$ as well: \cite{GULI} in this case no singularities are shown by $\partial_u \mathcal{S}$.

\section{Discussion of Numerical Results}

In order to study the transition to the insulating state at arbitrary $x$ we used the DMRG method. \cite{WHIDM} The algorithm has been customized \cite{ESPORT} for the convergence of not only the g.s., but also of several excited-state energies. This is important when the g.s. in a given sector of $N$ and $S^z_{\rm tot}$ is quasidegenerate and the procedure with only one target state could give unreliable results. The number of optimized DMRG states was fixed to (at least) $m=512$, and periodic boundary conditions have been used. The correlation functions on the g.s. are computed at the end of the finite-system iterations.

\subsection{Charge-gap closure}

We computed the charge gap at half-filling as
\begin{equation}\label{gap_eq}
    \Delta_{\rm c}=\frac{E(L+2)+E(L-2)-2E(L)}{2}\;,
\end{equation}
where $E(N)$ corresponds to the g.s. energy of the system evaluated as the lowest-energy state in the sector $S^z_{\rm tot}=0$ with $N$ electrons. Indeed, Eq. (\ref{gap_eq}) allows one to analyze the conducting properties of the system without affecting the magnetic ones.

\begin{figure}[!h]\label{grafico}
    \begin{center}
        \fbox {\includegraphics[height = 6 cm, width= 8 cm]{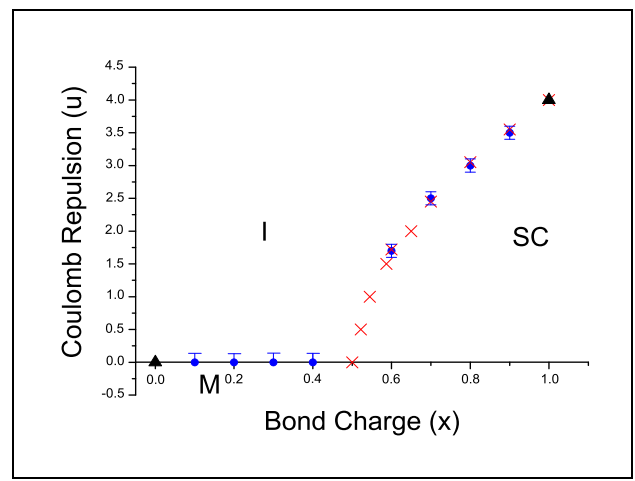}}
    \end{center}
    \caption{(Color online) The critical values for the opening of the charge gap in the $u,x$ plane, as obtained by the different methods: exact results (triangles), FSS analysis (dots), and singularities in $\partial_\alpha\mathcal{S}$, for $\alpha=u,x$ (crosses). M, I, and SC stand for metallic, insulating, and superconducting phases respectively.}
\end{figure}

\begin{figure}[!h]
    \begin{center}
        {\fbox{\includegraphics[height = 6 cm, width= 8 cm]{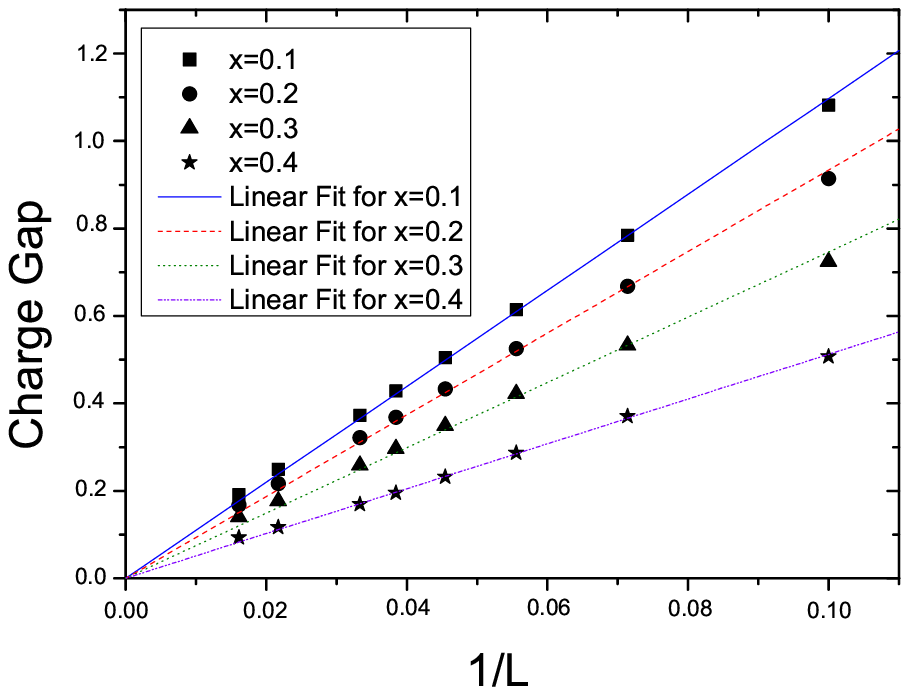}}}
    \end{center}
    \caption{(Color online) Finite-size scaling of the charge gap (\ref{gap_eq}) evaluated at $u=0$ for $x=0.1$, $0.2$, $0.3$, $0.4$, and $L = 10 - 62$ as a function of $1/L$.}\label{FSS}
\end{figure}

We fixed $x$ to different values and, for each of them, we evaluated the charge gap (\ref{gap_eq}) for several values of $u$ and $L$ in order to be able to perform the FSS analysis and to find the \emph{true} (that is, in the thermodynamic limit) critical value of $u$ for which the gap closes. We recall that, as $x$ grows, the system goes to a more symmetric and ordered phase, to the point that, at $x=1$, the g.s. becomes highly degenerate, and for $u\leq 4$ it also exhibits ODLRO, which implies strong long-ranged pair correlations. Accordingly, in the vicinity of that region the DMRG numerical simulation experiences some problems; the program precision becomes poorer, since the truncation procedure intrinsic to DMRG is expected to be less efficient when long-range correlations are present. Whereas when $u>u_{\rm c}(x)$ ---gapped phase--- numerical estimations are much more accurate: this allowed us to obtain ---with an acceptable error bar--- the critical points, as reported in Fig. \ref{grafico},
which summarizes our results, displaying how there are basically two regimes: $0\leq x < 0.5$ (i) and $0.5 < x \leq 1$ (ii). In (i), the transition takes place at $u_{\rm c}=0$: a small positive $u$ forces the system to become a charge insulator. Here the numerical simulation is not so demanding, and the results of FSS analysis of gap closure are quite accurate, as shown in Fig. \ref{FSS}.

\begin{figure}[!h]
    \begin{center}
        {\fbox{\includegraphics[height = 6 cm, width= 8 cm]{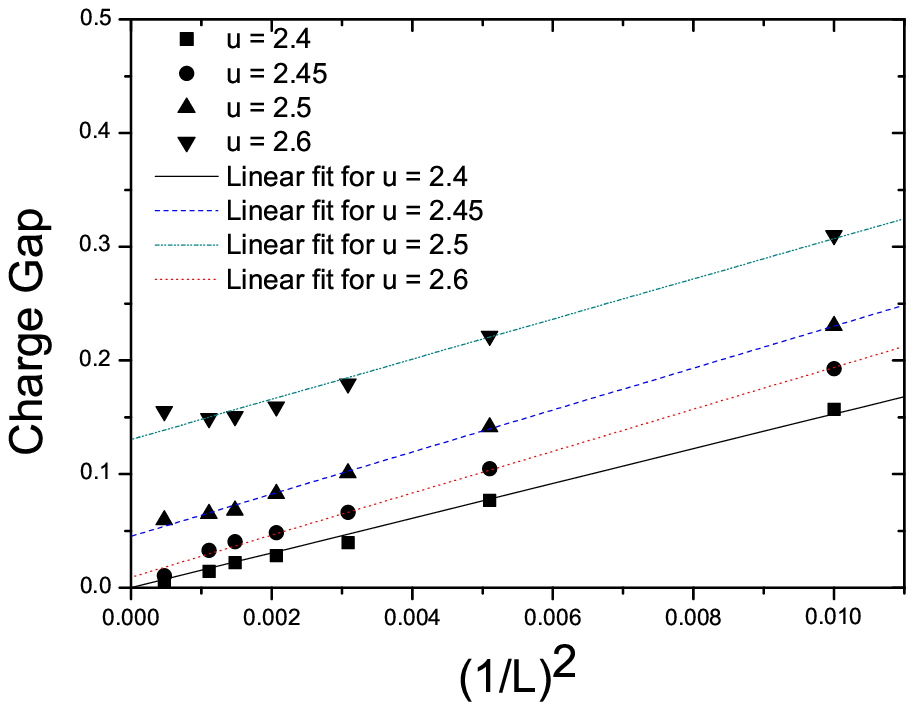}}}
    \end{center}
    \caption{(Color online) Finite-size scaling of the charge gap (\ref{gap_eq}) evaluated at $x=0.7$, $u=2.4$, $2.45$, $2.5$, $2.6$ and $L=10- 62$ as a function of $1/L^2$.} \label{NLFSS}
\end{figure}

On the other hand, in region (ii), $u_{\rm c}$ becomes different from zero and positive, increasing smoothly to match the analytical exact result $u_{\rm c}(1) = 4$. Correspondingly, a noninsulating phase is stable for positive values $u<u_{\rm c}$ of the Coulomb repulsion parameter. Interestingly, the behavior of the gap closure when entering such a phase changes; it is no longer of linear type, as reported in Fig. \ref{NLFSS}.

At $x=0.5$ the model Hamiltonian acquires extra symmetries, since the term interchanging singly and doubly occupied sites disappears. We observed that in its surrounding the numerical simulations, and in particular the FSS analysis, experienced greater difficulties, to the point that we were not able in this case to distinguish by this method whether or not the gap closes at finite $u_{\rm c}$ (up to $u_{\rm c}\lesssim 1$).

\subsection{Single-site entanglement}

As already pointed out, recent results have shown how at least some QPT's can be spotted by measures of entanglement. In particular, the transition to the insulating phase at $x=1$ is signaled as an algebraic divergence of $\partial_u \mathcal{S}$. \cite{AGMT} We then perform a study of the same transition at $x\neq 1$ through $\partial_\alpha \mathcal{S}$. Here $\mathcal{S}$ is a function of $n_d$ [see Eq. (\ref{le})]; an accurate estimate of $n_d$ is obtained from the correlation matrix $\langle n_i n_j\rangle$ ($n_i=\sum_\sigma n_{i\sigma}$) by averaging over its diagonal entries. This procedure compensates for the lack of a strict translational invariance due to the truncation approximation. Moreover, also finite-size effects are rather small, since $\mathcal{S}$ is related to single-site correlations and can be regarded as a thermodynamic quantity for $L\gg 1$; it turns out that its evaluation at $L=30$ gives already a good estimate of its large-$L$ behavior.

\begin{figure}[!h]
    \begin{center}
        \fbox{\includegraphics[height = 5.5 cm, width = 4 cm]{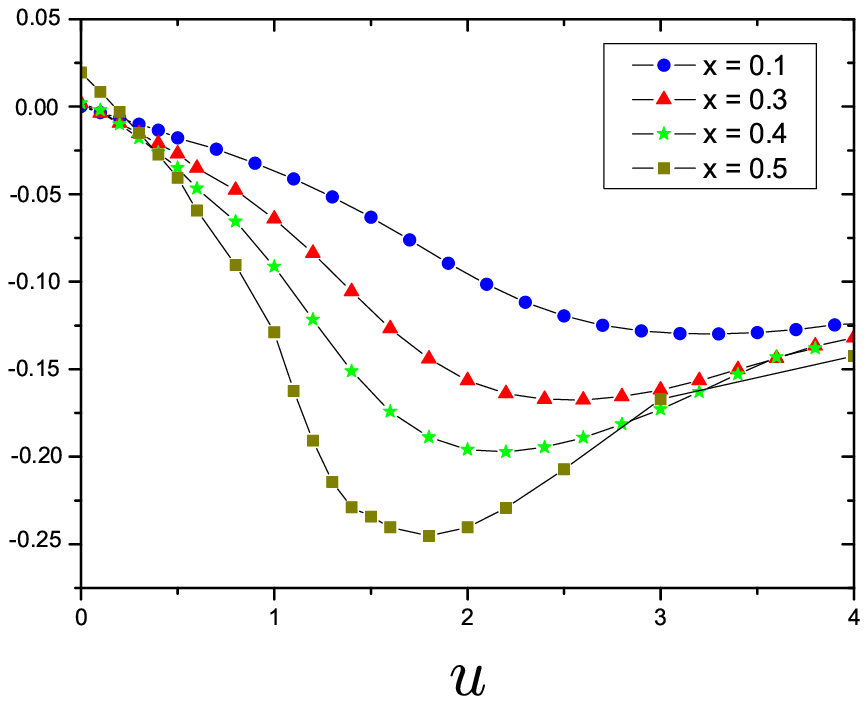}
        \includegraphics[height = 5.5 cm, width = 4cm]{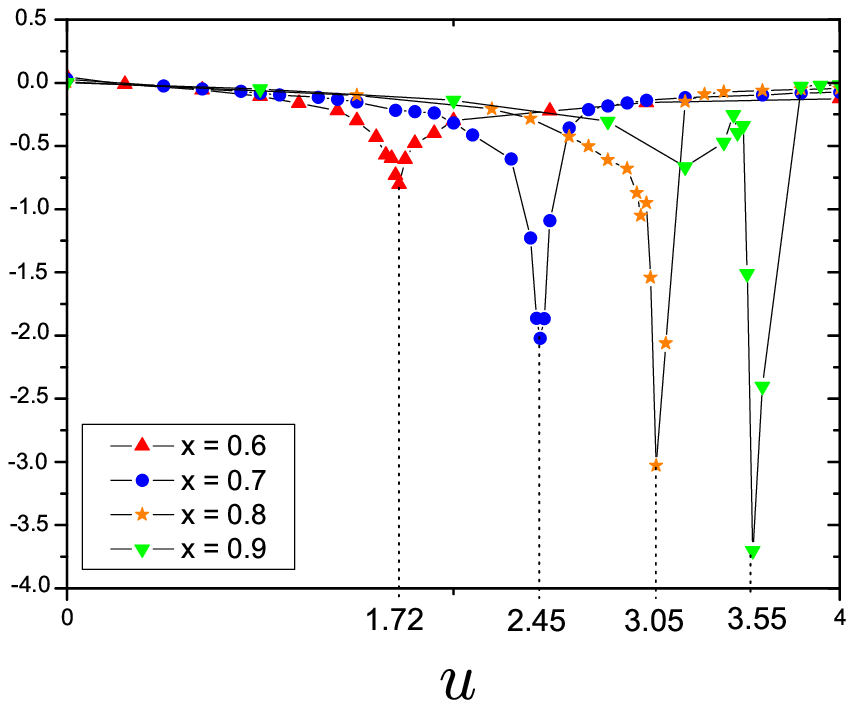}}
    \end{center}
    \caption{(Color online) $\partial_u \mathcal{S}$ as a function of $u$, evaluated at $L=30$  for (left) $x=0.1$, $0.3$, $0.4$, $0.5$ and (right) $x=0.6$, $0.7$, $0.8$, $0.9$.} \label{Dx7}
\end{figure}

The qualitative behavior of $\mathcal{S}$ for different $x$ is similar, being a decreasing monotone function of $u$ with a flex depending on $x$. Nevertheless, the behavior of its derivatives $\partial_\alpha \mathcal{S}$ changes drastically as $x\approx 0.5$, $u>0$. In Fig. \ref{Dx7} we report it for $\alpha=u$: a strong signature of the undergoing phase transition is seen whenever $x>0.5$, depicted as a sharp singularity at a critical point $u_{\rm c}'(x)$, which should be reminiscent of the divergent behavior of the same quantity in the thermodynamic limit. Remarkably, it turns out that $u_{\rm c}\equiv u_{\rm c}'$ within the whole range of $x$ values for which $u_{\rm c}$ can be evaluated through FSS (see Fig. \ref{grafico}). Moreover, for the reasons explained above, $u_{\rm c}'$ has negligible error bar, as opposite to $u_{\rm c}$.
Since ${\mathcal S}$ is easily worked out for arbitrary $u$, $x$, we may use the study of its derivatives for guessing $u_{\rm c}$, $x_c$ in those regions ($x\approx 0.5$, $u<1$) where the study of gap closure through FSS is hard. Indeed, in Fig. \ref{grafico} the points indicated just with crosses along the critical curve are obtained simply as singularities of $\partial_x \mathcal{S}$.

Notice that the divergence in $\partial_u \mathcal{S}$ disappears for $x\leq 0.5$. We recall that for $x<0.5$ we obtained $u_{\rm c}=0$; we then conjecture $u_{\rm c}'=0=u_{\rm c}$ also at $x=0.5$. We may infer that the different behavior in $\partial_u \mathcal{S}$ is related to the phase which opens below the insulating phase: a smooth behavior is found when such phase is metallic ($x\leq 0.5$), whereas a singular behavior appears when the phase changes (see below). This is signaled by the fact that also $\partial_x \mathcal{S}$ displays a singular behavior in correspondence of $x_{\rm c}$ whenever $u>0$. In terms of the continuum theory, one passes from a situation that is physically similar to the case $x=0$, where the gap is created by a \emph{marginally} relevant operator (as predicted by bosonization) and is characterized by an essential singularity, to a regime where the charge gap $\Delta_{\rm c} \propto (u-u_{\rm c})^\nu$ (with $\nu < \infty$) is generated by a truly relevant operator. In fact, according to the scaling analysis in Ref. \onlinecite{Campos_etal}, $\partial_\alpha \cal S$ should diverge with an exponent $\rho=2-2\nu$ whenever $\nu < 1$, and the numerical data in the right panel of Fig. \ref{Dx7} clearly show a finite value of $\rho$ and thus of $\nu$.

As a matter of fact, let us notice that the singularities found in $\partial_\alpha {\mathcal S}$ are a direct consequence of the behavior of $\partial_\alpha n_d$, since $n_d$ in the present case is the only nontrivial correlation contained in ${\mathcal S}$
[see Eq. (\ref{le})]. Hence the change which takes place along the critical curve is driven by $n_d$ as a function of $u$ or $x$.

\subsection{Pairing and spin-gap}

\begin{figure}[!h]
    \begin{center}
        \fbox {\includegraphics[height = 6 cm, width= 8 cm]{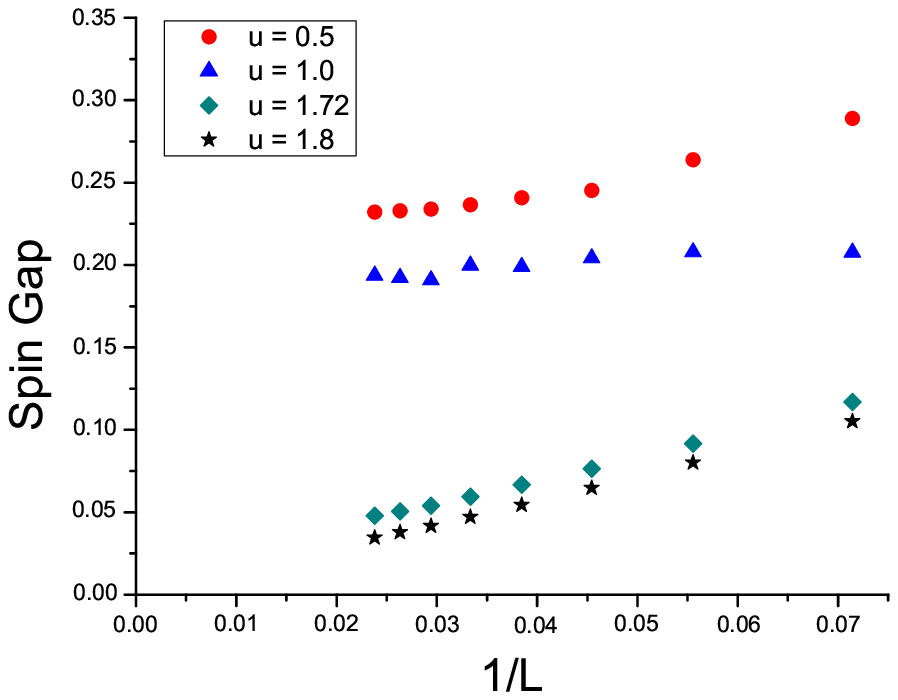}}
    \end{center}
    \caption{(Color online) Spin gap vs chain length at $x=0.6$ and various $u$ values.} \label{SG06}
\end{figure}

A study of the binding energy
\be
    E_{\rm b} \doteq\frac{1}{2}[E(L+2)+E(L)-2E(L+1)]
\ee
confirms that for $x>x_{\rm c}$, $u<u_{\rm c}$ a new phase opens, characterized by negative $E_{\rm b}$. This fact could correspond simply to phase separation, as well as to superconductivity, both phenomena having been observed in previous numerical studies by varying $x$, $u$, and $n$. \cite{AGAH,QUSH} In order to understand the nature of the new phase in our case, we have studied both singlet pair correlations $\langle\eta_i^\dagger \eta_j\rangle$ and the spin gap
\be
    \Delta_s= E(L,S^z_{\rm tot}=1)-E(L,S^z_{\rm tot}=0)
\ee
at $x=0.6$ at various $L$. The results support the superconducting (SC) nature of the phase. In fact, within such phase, the spin gap turns out to be open up to $u_{\rm c}$ (see Fig. \ref{SG06}).

\begin{figure}[!h]
    \begin{center}
        \fbox{{\includegraphics[height = 6 cm, width= 8 cm]{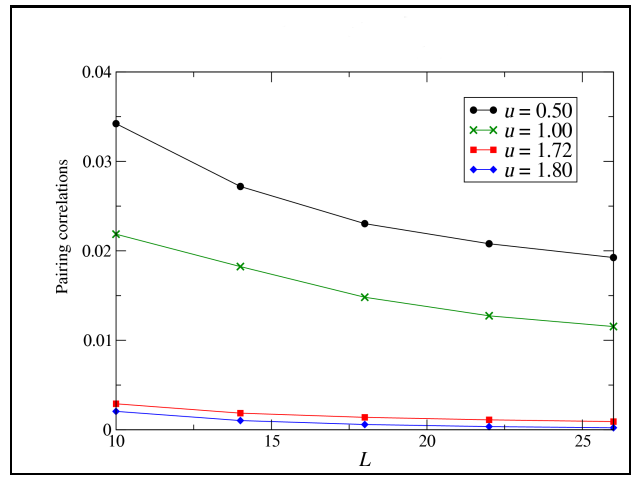}}}
    \end{center}
    \caption{(Color online) Pair correlations at half chain ($|i-j|=L/2$) vs chain length at $x=0.6$ and various $u$ values.}\label{PC06}
\end{figure}

Furthermore, in Fig. \ref{PC06} it is shown that as $u<u_{\rm c}$ the pair correlations do not vanish for $|i-j|\rightarrow\infty$, reaching an asymptotic value of the order of $n_d^2$, as in the exact case $x=1$.
A detailed study of the above phase would require an exhaustive investigation of spin gap and pair correlations for higher $x$ values as well. Such analysis is beyond the purposes of the present paper and is expected to become more demanding as $x$ becomes closer to $1$. Moreover, at $x=1$ the spin gap disappears. However, as far as the physics is concerned, since we do not expect that this would differ qualitatively from that observed at the specific value $x=0.6$, we resolved to call that phase SC. Notice also that in real materials it is more likely to have $x\approx 0.6$ than $x=1$.

\section{Conclusions}

In summary, we have studied the role played by the bond-charge interaction in the opening of the insulating phase in the Hirsch model. The critical curve ($u_{\rm c}$, $x_{\rm c}$) is accurately determined through the composite use of FSS analysis of charge-gap closure and of singularities in $\mathcal{S}$; the curve qualitatively agrees with results obtained within the slave-boson approximation. \cite{BUL} The point $x=0.5$ appears to play a key role. In particular, the bosonization description, indicating $u_{\rm c}=0$, holds up to $x\approx 0.5$, whereas the rich physics characteristic of the integrable (and highly degenerate) case $x=1$ in fact qualitatively survives down to $x\approx 0.5$. The region below $u_{\rm c}$ changes from metallic to SC again at $x\approx 0.5$. In correspondence, the FSS analysis of gap closure shows that the dependence of $\Delta_{\rm c}$ on $1/L$ at the critical point passes from linear to nonlinear, which fact may be interpreted within conformal field theory \cite{HENKEL} as a signature of vanishing charge velocity $v_{\rm c}$. This possibility is supported both by a direct calculation at $u=2.45$, $x=0.70$, where we found $v_{\rm c} \simeq 10^{-2}$, and by Fig. \ref{FSS} in which the line slope decreases when approaching $x=0.5$ from below. Note that such a scenario requires that the central charge along the critical curve be $c=1$, for both the charge and spin degrees of freedom. Using the formula in Ref. \onlinecite{CC} for the so-called block entropy  we have checked that this is indeed the case. In terms of the quantum-classical mapping \cite{Sachdev} the $1/L^2$ behavior suggests that the dynamical exponent is $2$ for $x>0.5$, which might be interpreted as the system approaching a parabolic point in the band.

The procedure of using the single-site entanglement as a witness of ongoing QPT's is quite convenient if considered from a time-consuming perspective, since it does not require a thorough FSS analysis (at least in this context). Moreover, once it is found that a particular transition (like the SC-insulator transition) is captured by such procedure, this could be used beyond the one-dimensional case, by simple implementation of exact-diagonalization techniques on small two-dimensional clusters.

Even within the one-dimensional case, moving away from half-filling should provide other interesting physics, since particle-hole invariance for $x\neq 1$ is lost. It is expected that within the hole-doped region higher values of $u_{\rm c}$ for the SC phase could be achieved; \cite{QUSH} though such a QPT is of a different nature at $n\neq 1$, it should be captured by appropriate measures of multipartite entanglement. \cite{AGMT}

\begin{acknowledgments}
The authors gratefully acknowledge interesting discussions with F. Dolcini, E. Ercolessi, G. Morandi, P. Giorda, and M. Roncaglia, as well as useful comments by L. Arrachea. This work was partially supported by the Italian MIUR (Grant Nos. 2002024522\_001 and
2003029498\_013).
\end{acknowledgments}

\end{document}